\documentclass[pss]{wiley2sp}
\usepackage{amsmath}
\usepackage{color}
\usepackage{graphicx}
\usepackage{cancel}
\usepackage{amsfonts}

\def  \bsig    {\mbox{\boldmath$\sigma$}}

\def    \hp       {\hat{p}}
\DeclareMathOperator{\sech}{sech}
\DeclareMathOperator{\sgn}{sgn}

\begin{document}

\title{Indirect interaction of magnetic domain walls}

\titlerunning{Indirect interaction of domain walls}

\author{%
  N. Sedlmayr\textsuperscript{\Ast,\textsf{\bfseries 1}},
  V. K. Dugaev\textsuperscript{\textsf{\bfseries 2,3}},
  M. Inglot\textsuperscript{\textsf{\bfseries 2}},
  J. Berakdar\textsuperscript{\textsf{\bfseries 4}}}

\authorrunning{N. Sedlmayr et al.}

\mail{e-mail
  \textsf{sedlmayr@physik.uni-kl.de}, Phone:
  +49-631-205 2393}

\institute{%
  \textsuperscript{1}\,
  Department of Physics, University of Kaiserslautern, 67663 Kaiserslautern, Germany\\
  \textsuperscript{2}\,
  Department of Physics, Rzesz\'ow University of Technology,
  Al.~Powsta\'nc\'ow Warszawy 6, 35-959 Rzesz\'ow, Poland\\
  \textsuperscript{3}\,
  Department of Physics and CFIF, Instituto Superior T\'ecnico,
  TU Lisbon, Av.~Rovisco Pais, 1049-001 Lisbon, Portugal\\
  \textsuperscript{4}\,
  Martin-Luther-Universit\"at Halle-Wittenberg, Heinrich-Damerow-Str. 4,
  06120 Halle, Germany}

%\received{XXXX, revised XXXX, accepted XXXX} % do not change, will be filled in by the publisher
%\published{XXXX} % do not change, will be filled in by the publisher

\keywords{Domain walls, exchange interaction, RKKY, spintronics}

\abstract{
We calculate the electron-mediated exchange interaction between two domain
walls in magnetic wires. This corresponds to the equilibrium regime and,
therefore, the interaction can be additionally controlled by an electric current.
The exchange interaction is long ranged and oscillates as a function
of the distance between the walls. It also depends oscillatory on the
polarization angle of the walls, having the maximum value for collinear
polarization.}
%
% This is a macro for the typesetting of two-column text in an
% abstract. It will typeset the two arguments in \abstcol{}{} as the
% left and right column inside the abstract box. At the
% columnbreak there will be always a columnbreak (\par), so both
% columns start with a new paragraph. No automatic column height
% balancing is done.
%
% If used with a \titlefigure it will silently output both
% parameters as consecutive paragraphs.
%
% The macro is defined exclusively inside the argument of \abstract{};
% if used outside it will raise an error.
%
% Usage: \abstcol{<left column>}{<right column>}
%\abstcol{%}{}}

% The class file requires the standard graphicx Latex package. See the 'LaTeX
% standard graphics and color packages documentation' for more information at
% <http://tug.ctan.org/tex-archive/macros/latex/required/graphics/grfguide.pdf>.
%
% Accepted figure file formats depend on which LaTeX flavour is used.
% Classic LaTeX is always able to use Encapsulated Postscript (EPS);
% PDFLaTeX can't use this but accepts PDF, JPG, PNG, and GIF formats.
%
% See examples for implementing graphics in floating figure environments later in this file.
% If \titlefigure is given, it takes as its mandatory parameter the
% name (without extension) of some figure file.
%\titlefigure[height=3.1cm]{empty2w}
%\titlefigurecaption{%
%  This is the caption of the \emph{optional} abstract figure. If
%  there is no abstract figure here, the abstract text should be divided into both columns.}

\maketitle   % please do not remove

\section{Introduction}

Magnetic domain walls (DWs) have attracted a lot of attention as important elements
of new magnetoelectronic devices \cite{allwood05,chappert07,klaui08}.
In particular, it was demonstrated recently that they can be used in a
new type of memory device (the racetrack memory) effectively controlled by an electric
current \cite{parkin08,parkin10}. On the other hand, the DW
can be viewed as a kind of local imperfection in an ordered magnetic
system, like an impurity or defect. The substantial difference is that DWs can
move and therefore they can be relatively easily put into motion by an external
field or electric current, and also by an interaction between them.

A great amount of theoretical and experimental work
was dedicated to studying the resistance of
DWs, the current-induced spin torque, the dynamics of DW motion, and other effects
related to a single DW strongly coupled to the electron system
\cite{Marrows2005,klaui08,sedlmayr11a}.
When the density of DWs
increases, it is important to include into these considerations the effects of their
interaction. It has already been demonstrated \cite{sedlmayr09,sedlmayr20101419,sedlmayrpssb,sedlmayr11}
that an electric current in a magnetic
wire with DWs influences the DW interaction, so that by using a current the DW coupling
can be controlled, and the dynamics of strongly coupled DWs can be affected.

In this work we consider in detail the indirect exchange coupling between the DWs in
equilibrium. Essentially, the analysis of such interaction is the first necessary step
to understand the basic mechanisms of DW interactions.

\section{Model}

Let us consider the following Hamiltonian, describing an electron gas coupled by exchange energy $M$ to
a textured magnetization which changes its orientation at each DW:
\begin{eqnarray}
\label{1}
H=-\frac{\Delta }{2m}-M\bsig \cdot {\bf n}({\bf r}),
\end{eqnarray}
where the unit vector ${\bf n}({\bf r})$ determines the magnetization orientation, $m$ is the carrier effective mass, 
and we take units with $\hbar=1$.
We assume the ferromagnetic wire (or ribbon) to be orientated along the $x$ axis and
consider two DWs, labeled as 1 and 2.
For definiteness, we also assume that at $x\to - \infty $ and $x\to +\infty $
the magnetization is along unit vector ${\bf n_0}=(0,\, 0,\, 1)$,
and within each of the DWs the vector ${\bf n}$ is rotated
by an angle $\pi $ around the unit vector ${\bf t}_i$, where the index $i=1,2$ refers
to DW 1 or 2. We focus here on the case of transverse DWs, the results for a vortex DW should look qualitatively similar.
Thus, the dependence of ${\bf n}(x)$ within DWs 1 and 2 is described as
\begin{eqnarray}
\label{2}
&&{\bf n}_1(x)=e^{i\varphi _1\, {\bf t}_1\cdot {\bf L}}\, {\bf n}_0,
\hskip0.5cm
{\bf n}_2(x)=-e^{i\varphi _2\, {\bf t}_2\cdot {\bf L}}\, {\bf n}_0,
\end{eqnarray}
where ${\bf L}$ is the matrix of moment $L=1$
and $\varphi _i(x)$ changes from 0 to $\pi $ when $x$ crosses the $i$-th DW.
Correspondingly, the transformation of the spinor wave function $\psi $ to
the local frame with the magnetization along the axis $z$ is
$\psi \to \hat{T}\psi =e^{\frac{i\varphi }2\, {\bf t}_i\cdot \bsig }\, \psi $.
The location $x_i$ and the vector ${\bf t}_i$ determine the DW state.
Note that for the electron motion the DW can be always considered as static
even for DWs moving along the wire.

We use this transformation to the local frame \cite{kor,tatara,dugaev1},
in which the vector ${\bf n}_i$ of each DW is oriented in the same direction
along the global axis $z$. After this transformation the electron gas
is in the homogeneous magnetization $M$ but there appears the gauge potential
related to the local transformation,
$A_i(x)=i(\varphi '_i/2)\, ({\bf t}_i\cdot \bsig )$.

We assume the DW width $\lambda$ to be much larger than the electron wavelength
at the Fermi surface, $k_F\lambda\gg 1$,
which is the typical condition for metallic ferromagnets. Then the transformed Hamiltonian is (summing over $i$)
\begin{eqnarray}
H&=&\frac{k_y^2+k_z^2}{2m}
-\frac{\hp_x^2}{2m}-M\sigma _z
-\left[\frac{\beta_i}{2}  {\bf t}_i\cdot \bsig \, \hp_x+\textrm{h.c.}\right] ,\label{4}
\end{eqnarray}
where $\beta_i (x)=\varphi'(x-x_i)/2m$, $\hp_x=-i\partial/\partial x$
and $x_i$ is the point where the $i$-th DW is located.

The exchange interaction energy can be found using a RKKY approach
with the DW-induced perturbation localized in the vicinity of the points
$x_1=0$ and $x_2=R$. Using Eq.~(3) we find
\begin{eqnarray}
\nonumber
E_{int}
=\sigma_{cs}\, {\rm Re}\, {\rm Tr}
\int \frac{d^2k}{(2\pi )^2}\, \frac{d\varepsilon }{2\pi }\;
dx'\, dx''\, \beta _1(x')\, \beta _2(x'')\,
\nonumber \\ \times
({\bf t}_1\cdot \bsig )\,
\frac{dG_{k\varepsilon }(x'-x'')}{dx'}\,
({\bf t}_2\cdot \bsig )\,
\frac{dG_{k\varepsilon }(x''-x')}{dx''} ,
\label{7}
\end{eqnarray}
where $\sigma_{cs}$ is the cross-section of the DW and
\begin{eqnarray}
\label{8a}
G_{k\varepsilon }(x)
=\int _{-\infty }^{\infty }\frac{dk_x}{2\pi }\, e^{ik_xx}\, {\rm diag }
\left(G_{\vec{k}\varepsilon \uparrow }\, ,G_{\vec{k}\varepsilon \downarrow }
\right)
\end{eqnarray}
is the Green function of electrons in a homogeneous magnetization field: $G^{-1}_{\vec{k}\varepsilon \sigma }=\varepsilon-\varepsilon_{k\sigma}-k_x^2/2m+\mu+i\delta\sgn(\varepsilon)$, where $\mu$ is the chemical potential.
We also denoted $\varepsilon _k=(k_y^2+k_z^2)/2m$ and
$\varepsilon_{k\uparrow ,\downarrow }=\varepsilon _k\mp M$.
Now we take
$k_\sigma =+\sqrt{2m(\varepsilon -\varepsilon _{k\sigma }+\mu )
+i\delta \, {\rm sgn}\, \varepsilon }$.
Then defining $\xi_{k\sigma}=\varepsilon _{k\sigma }-\mu$ we have the conditions that if $\varepsilon-\xi_{k\sigma}>0$ and $\varepsilon<0$ then $k_\sigma$ lies in the lower half-plane, otherwise it lies in the upper half plane. This allows us to directly calculate $G_{k\varepsilon }(x)$.

Calculating the Green functions and their derivatives
in coordinate representation and
substituting them into Eqn.~(\ref{7}) we obtain
\begin{eqnarray}
\label{8b}
E_{int}
&=&-\sigma_{cs}m^2\, {\rm Re}\, {\rm Tr}\int \frac{d^2k}{(2\pi )^2}\,
\frac{d\varepsilon }{2\pi }\,
dx'\, dx''\,
\nonumber\\
&&\times \beta _1(x')\, \beta _2(x'')\,
({\bf t}_1\cdot \bsig) \,
\mathbf{P}\,
({\bf t}_2\cdot \bsig) \,
\mathbf{P} ,
\end{eqnarray}
where we denoted
\begin{eqnarray}
\label{8c}
\mathbf{P}&=&{\rm diag }\big(f_{k\uparrow}
e^{ik_\uparrow (x''-x')}%\\&&
+(1-f_{k\uparrow})e^{ik_\uparrow (x''-x')\sgn(\varepsilon)},
\nonumber\\&&  f_{k\downarrow}e^{-ik_\downarrow (x'-x'')}+(1-f_{k\downarrow})
e^{-ik_\downarrow (x'-x'')\sgn(\varepsilon)}\big)\hskip0.2cm%\nonumber
\end{eqnarray}
and $f_{k\sigma }\equiv f(\varepsilon _{k\sigma })$ is the Fermi-Dirac function
at $T\to 0$.
The traces over the matrices can be performed immediately. Assuming that the first
DW points into the $y$-axis, then we can take
${\bf t}_1=\hat{x}$ and ${\bf t}_2=\hat{x}\cos\theta +\hat{y}\sin \theta $, so
that $\theta$ is the angle between the two domain wall polarizations.
At $T=0$ this yields
\begin{eqnarray}
\label{10}
&&{\rm Tr}\, ({\bf t}_1\cdot\bsig)\, \mathbf{P} \,
({\bf t}_2\cdot \bsig)\,\mathbf{P}
=2\cos(\theta)
\nonumber \\
&&\qquad\times
\big[f_{k\uparrow}e^{-i(x'-x'')(k_\uparrow+k_\downarrow)}
+e^{-i(x'-x'')(k_\uparrow+k_\downarrow{\rm sgn}\, \varepsilon) }
\nonumber \\
&&\qquad+(1-f_{k\downarrow})e^{-i(x'-x'')(k_\uparrow
+k_\downarrow) {\rm sgn}\, \varepsilon}\big] .
\end{eqnarray}
For the $\beta $ function we use
$\beta_1(x)=\sech(x/L)/2mL$, then
we can calculate the integrals over $dx'$ and $dx''$ using
\begin{eqnarray}
\label{beta1}
&&\int dx'\, \beta_i (x')\, e^{\mp ikx'}
=\frac{\pi}{2m}\sech \frac{\pi Lk}2\, e^{\pm ikR_i}
%\\
%\label{beta2}
%&&\int dx''\, \beta_2 (x'')\, e^{ikx''}
%=\frac{\pi}{2m}\sech \frac{\pi Lk}2\, e^{ikR}.
\end{eqnarray}
for $i={1,2}$ and $R_1=0$, $R_2=R$. Now we can switch the $k$-integrals to polar coordinates and perform the angular integral,
and then substituting $\varepsilon'=k^2/2m$ and
rescaling $\epsilon=\varepsilon-\varepsilon'+\mu$ we obtain
\begin{eqnarray}
E_{int}(\theta,R)
=-\frac{m\sigma_{cs}\cos \theta }{32}\, {\rm Re}\bigg[
\int_0^{\infty} d\varepsilon' \int_{M}^\infty
d\epsilon \, e^{iRk_1}
\nonumber\\ \times
\sech^2 \frac{\pi L(k_1+k_2)}2\,
\left( e^{iRk_2}
+e^{iRk_2\sgn(\epsilon+\varepsilon'-\mu)}\right)\\ +2M
\int_{M}^\infty
d\epsilon \, e^{iR(k_1+k_2)}
\sech^2 \frac{\pi L(k_1+k_2)}2\bigg] ,\nonumber
\end{eqnarray}
where $k_{1,2} =\sqrt{2m(\epsilon\pm M)}$.

\begin{figure}
\includegraphics*[width=0.42\textwidth]{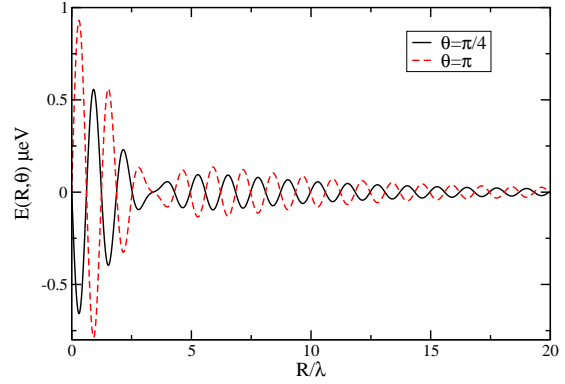}
\caption[Energy]{The interaction energy as a function of inter DW distance $R$, for  $\lambda=10\lambda_F$ and two different angles between the DWs. See text for details.}
\label{energy3}
\end{figure}

\begin{figure}
\includegraphics*[width=0.42\textwidth]{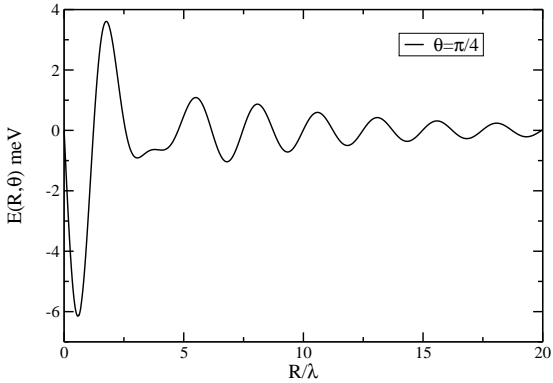}
\caption[Energy]{The interaction energy as a function of inter DW distance $R$,
for $\theta=\pi/4$ and $\lambda=5\lambda_F$. See text for details.}\label{energy2}
\end{figure}

The dependence of the exchange interaction on the distance for the two
polarization angles $\theta $ is shown in Fig.~\ref{energy3}.
Here we use the following parameters: the Fermi wavelength
$\lambda_F=0.367$~nm, $M=36$~meV, $\sigma_{cs}=20\times 20$~nm${}^2$, and $\lambda=10\lambda_F$.
The magnitude of the interaction depends strongly and non-monotonically on the DW
width, $\lambda$, and the magnetization strength, $M$. They are related
by the strengths of the anisotropy and the exchange energy in the material.
%With a fixed anisotropy the limit of vanishing exchange, to the non-magnetic case,
%should be treated with caution, as it can break the approximations used to treat
%the gauge potential.
Halving the width of the DW to $\lambda=5\lambda_F$, we already
see a much larger effect, see Fig.~\ref{energy2}.

The obtained results show that the interaction of two DWs is long-ranged and strongly
oscillating.
The particular behaviour around $R\approx4\lambda$ is caused by the change of sign of the envelope function of the energy. I.e.~the envelope function is also not a monotonic function of $R$.
But the most important is that the DW interaction depends on both the
distance between the walls and the DW polarization determined by the angle
$\theta$. It means that if we put DWs
located at a certain distance $R$, which corresponds to the energy-favorable
collinear mutual polarization of DWs,
then there is another location very close in $R$ with the anti-collinear orientation of DWs,
with almost the same energy. Considering the energy of the system as a function
of distance $R$, we find the correspondence to a classical ``particle'' in the
oscillating potential profile, so that the neighboring positions of this particle
in the minima of the potential
describe the up and down states of one of the DWs with respect to the other one.
As the amplitude of interaction increases with decreasing $R$ at small
distances (see Figs.~1 and 2), the DWs are effectively attracted to each other.

One can assume that one of the DWs is not moving (e.g.,  due to  pinning). Let us assume that the
classical particle representing the other DW is located in one of the potential minima.
If the distance between the minima is small, the particle can tunnel through the barrier,
so that the other DW can be presented as a delocalized quantum particle.
The Hamiltonian which describes it, has the form
$H=t\sum _i(c^\dag _{i\sigma }\sigma ^x_{\sigma \sigma '}c_{i+1,\sigma '}+h.c.)$
where spin up and down states correspond to the collinear and anti-collinear
orientations of the second DW with respect to the first one.

\begin{figure}
\includegraphics*[width=0.5\textwidth]{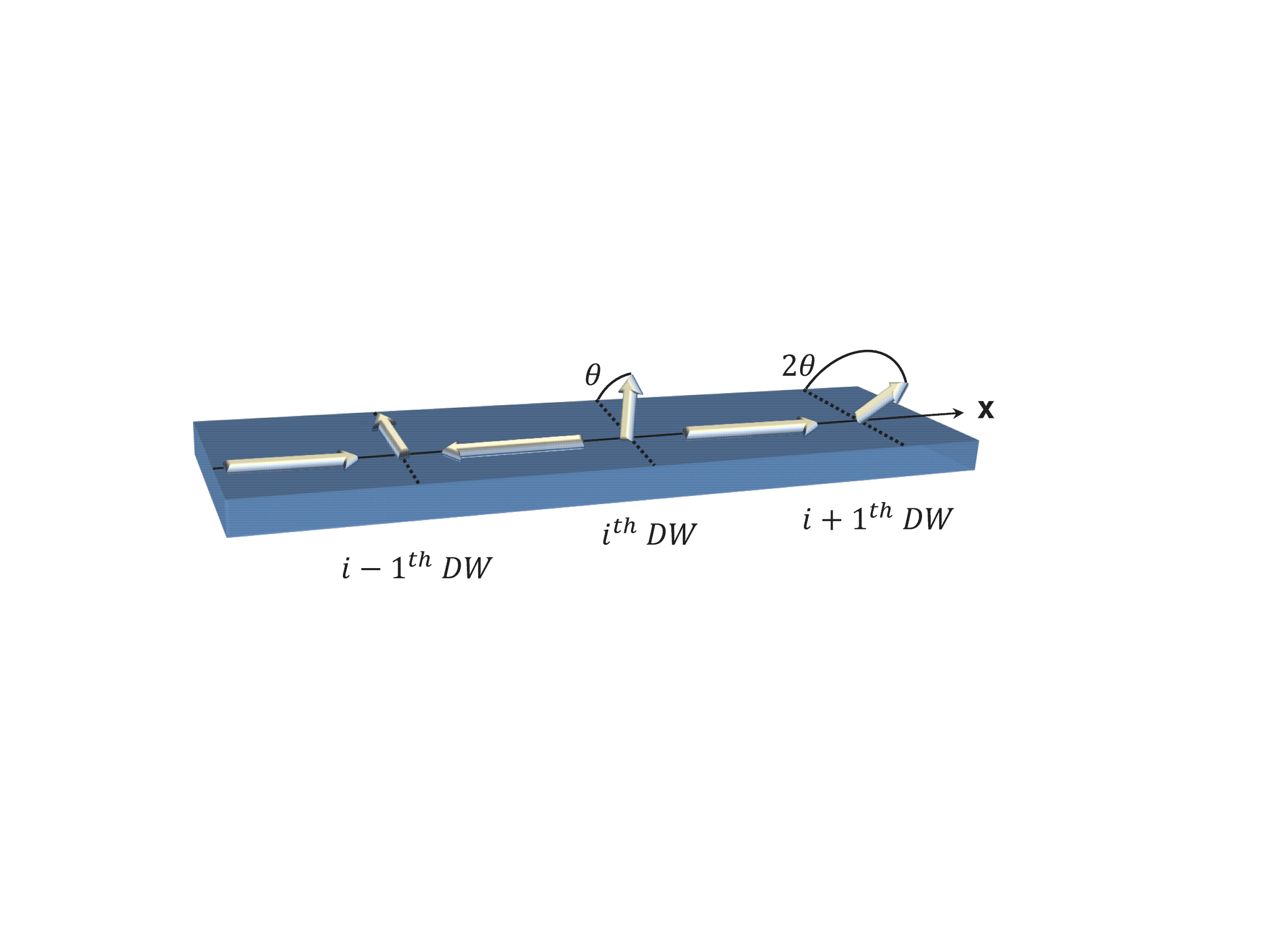}
\caption[Sppiral]{A schematic of the spiral structure of the DWs for $J_2<0$ and $|2J_2|>|J_1|$, the configuration of three of the DWs is shown. $\theta=\theta_{i+1}-\theta_i=\cos^{-1}[-J_1/2J_2]$. The DW orientation, defined at the centre of the DW, and the bulk magnetization are shown.}\label{spiral}
\end{figure}

Classically, if we consider a series of DWs pinned at a specific distance from each other,
$R^*$, inside a wire then the total energy of the system is given by an XY-model
\begin{eqnarray}
\label{13}
E=-\sum_i\left[J_1\cos(\theta_{i+1}-\theta_i)
+J_2\cos(\theta_{i+2}-\theta_i)\right] ,
\end{eqnarray}
where only the nearest neighbour and next nearest neighbour interactions are included.
Then we have
$J_1=-E_{int}(0,R^*)$ and
%,\hskip0.5cm
$J_2=-E_{int}(0,2R^*)$.
%\end{eqnarray}
In general we can have $J_1$ and $J_2$ as either negative or positive and either
$|J_1|>|J_2|$ or $|J_2|>|J_1|$. There are several possible set-ups admitting a simple
solution.
If $|J_1|\gg|J_2|$ then we have either an antiferromagnet:
$\theta_{i+1}-\theta_i=\pi$ (for $J_1<0$), or a ferromagnetic arrangement
$\theta_{i+1}-\theta_i=0$ (for $J_1>0$). If we take $J_1\to 0$ then we get two
sublattices with either AFM or FM arrangements depending on the sign of $J_2$.
If we have $2J_2<J_1$ and $2J_2<-J_1$, i.e. $J_2<0$ and $|2J_2|>|J_1|$, then the
model is minimized by $\cos[\theta_{i+1}-\theta_i]=-J_1/2J_2$. In this case
we obtain a spiral structure of the DW orientations through the wire, see figurte \ref{spiral}.
Experimentally, similar spiral structures have been already observed \cite{PhysRevLett.93.117205,PhysRevLett.103.157201}.

\section{Conclusion}

In summary we have found that there exists an RKKY-like electron mediated interaction
between DWs in a ferromagnetic sample that is long-range and oscillating. This interaction
remains, \emph{independent} of any current flowing through the system, in addition to previously
found results for current mediated interactions in nanowires.

\begin{acknowledgement}
This work is partly supported by FCT Grant No.~PTDC/FIS/70843/2006 in Portugal, by the DFG contract BE
2161/5-1, and by the Graduate School of MAINZ (MATCOR).
J.B.~acknowledges financial support by Stanford Pulse Institute and
Stanford Institute for Material and Energy Science.
\end{acknowledgement}

\providecommand{\WileyBibTextsc}{}
\let\textsc\WileyBibTextsc
\providecommand{\othercit}{}
\providecommand{\jr}[1]{#1}
\providecommand{\etal}{~et~al.}

%\bibliographystyle{pss}
%\bibliography{referencesdwrkky}

\end{document}